# Modelling and Classification of Fairness Patterns for Designing Sustainable Information Systems

Christophe Ponsard[1], Bérengère Nihoul[1], Mounir Touzani[2]

[1]CETIC - Research Centre, Gosselies, Belgique, {christophe.ponsard, berengere.nihoul}@cetic.be
[2]Independant researcher, Toulouse, France, mounir.touzani@inrae.fr

**ABSTRACT.** Designing sustainable systems involves complex interactions between environmental resources, social impacts, and economic issues. In a constrained world, the challenge is to achieve a balanced design across those dimensions while avoiding several barriers to adoption. This paper explores the concept of fairness in sociotechnical system design, including its information system component. It is based on a reference sustainability meta-model capturing the concepts of value, assumption, regulation, metric and task. Starting from a set of published cases, different fairness patterns were identified and structured in a library enabling the application of strategies for adoption, anticipation, distributive justice, and transparency. They were generalised and documented using an existing sustainability template. An extension to the initial meta-model is also proposed to identify and reason on assumptions and barriers to reach the desired values. Finally, the validation of our work is discussed using two case studies, respectively addressing the fairness to manage the COVID-19 crisis and the medico-social follow-up of childhood.

## Introduction

Sustainability is a broad and multidimensional concept. The United Nations Brundtland Report states : « Sustainable development is a process of change in which the exploitation of resources, the direction of investments, the orientation of technological development ; and institutional change are all in harmony and enhance both current and future potential to meet human needs and aspirations » [Uni87]. This wise exploitation of current resources and the future of all humanity has an impact on how to design our systems. Information systems and software are also involved : they are both the source of problems because of their environmental footprint, but also contribute to the implementation of solutions [CP15]. Defining the sustainability of software is just as difficult and is still being debated, particularly from the point of view of non-functional requirements and emerging properties [Ven+14].

Sustainability involves a notion of fairness in the resource allocation process among actors involved in the system. This is reflected by Mahatma Gandhi's quote : « There is enough for everybody's need, but not enough for anybody's greed ». It also means having to impose constraints on the way systems are designed to be balanced, not necessarily aiming for pure formal equality which is not achievable in practice, but to achieve an overall equilibrum based on a fundamental value such as solidarity. For example, the healthy help the sick, the adult takes care of young/older, because everybody is likely to go through those various roles. This approach is also highlighted by social or distributive justice [Raw71][Mai13].

Defining fairness is just as difficult as defining sustainability, because different cultures have different interpretations of this notion. Each society defines what is fair or unfair and produces understandable rules and standards. Essentially, it seems fair that those who deserve things should get them and that those who do not deserve them should not receive them [Wal16]. Beyond the notion of equality with regard to the possession of social goods, complementary approaches advocate the equality of »capabilities, i.e. the effective possibilities of carrying out certain acts [Sen09]. The term »substantial freedom « is also used for this concept.

This work is based on the notion of a Socio-Technical System (or system in the rest of this work). Such a system results from the combination of technical, human or natural elements. Indeed, sustainability calls for a global analysis combining in a transverse way hardware, software, personal and community components in order to take into account elements relating to social structures, roles and regulations. The information system (IS) is an integral part of the system under consideration and itself comprises a technical component (software, hardware, network as well as processes) and a social component (organisation and people linked to the IS) [Pic12]. However, the system we are considering here goes far beyond the IS to avoid limiting ourselves to very technical definitions of fairness, for example, in terms of infinite sequences of states which must always see the achievement of a wished property after a sufficiently long time [Fra86]. Here we are in a perspective where it is imperative to also grasp the social, environmental, economic and personal dimensions [Pen+14].

In this context, our objective is to propose a tool-based methodology to help identify and manage fairness requirements in the context of socio-technical systems, and more specifically those concerning the IS to be put in place to serve an equitable socio-technical system. The approach adopted is based on the development of a knowledge base that can be easily structured, reused and mobilised in order to analyse and arbitrate decisions on the organisation of systems with regard to their desired fairness properties and within a more general framework of sustainability. These decisions concern global aspects of the system but also make it possible to identify more specific requirements for the implementation of an IS at the service of these properties. The focus is not on optimising the sustainable and fair design of the IS component alone but to make IS related requirements will emerge from the analyses carried out. Our work is structured around a consistent base of interesting case studies from the point of view of fairness. These cases come from our own research or have been reported in the literature by other teams active in this field.

**This article builds on our initial work [PNT21] and is the English translation of a paper in a French journal [Chr24]**. This work demonstrated the feasibility of a modelling-based approach to reasoning about the problem of fairness. It had sketched out an initial structuring in the form of patterns and had carried out a limited validation. The basis of this work remains similar and relies on an fairness reference model [Kie+20] to define a reasoning framework based on a meta-model that can be used for sustainability. It also continues to rely on graphical representations already in use in this field [PF13]. This work extends our initial results in the following way :

— it includes a more systematic and detailed review of the reference frameworks used.
— it relies on a much broader base (doubled) of case studies used to identify and document our patterns.
— our patterns are better structured and presented in the form of a richer and more detailed library while remaining based on a model of sustainability patterns [RR13].
— in addition to the management of the COVID19 pandemic, a validation was carried out on a prenatal and early childhood medical monitoring system where many questions of fairness of access to care are present.
— a broader discussion was held, in particular on the possibilities and benefits of extending the ap-

proach to sustainability analysis, or even proposing a different approach to non-functional requirements.

The paper is structured as follows : Section 1 provides a review of the state of the art and practices, giving existing foundations and published case studies on which to base our work. Section 2 describes the methodological approach used, including some motivated extensions of existing notations. It also illustrates the pattern discovery process on a case from the hospital domain. Section 3 presents the structure and representations used for our fairness pattern library. In Section 4, we present our validation cases composed of the COVID-19 crisis management and a prenatal/early childhood health monitoring system. Section 5 discusses our catalogue and its validation, in particular relating to modelling, methodology, completeness and generalisability aspects. Finally, we conclude by identifying avenues for further research.

## 1. State of the Art and of Practices

Before describing our approach, this section describes both the state of the art and ot the current practices. We first present some theoretical foundations. Next, we detail two existing approaches used in a combined and enriched manner. Finally, we identify case studies from the literature that have informed our approach to structuring fair design knowledge. With the exception of the last section, the material presented here is not specific to fairness but applicable to a knowledge acquisition approach to sustainability.

### 1.1. *Theoretical Foundations for the Sustainability of Socio-Technical Systems*

As a prelude, let us recall some findings reported in a systematic review of the literature carried out on the basis of more than 180 articles between 1990 and 2020 relating to the foundations of the concepts of sustainability and change in socio-technical systems [Sav+19]. This review is organised around three fundamental axes which will also be the structuring principles for the rest of our work :

— The (« WHY ») focuses on the goals of sustainability through a socio-technical approach. It ties in with the arguments raised in the introduction on the balance between the various dimensions of economic performance, social inclusion (including fairness) and environmental resilience. The need to reason in terms of socio-technical systems is also discussed in the light of the tensions and unanticipated consequences associated with the introduction of technology, whether in terms of generating inequality or environmental degradation. This highlights a more concrete notion of value. Innovation and the aspiration to progress are also identified as stimuli for removing barriers and making systems more sustainable.

— The (« WHAT ») focuses on the multiple interpretations of what needs to be made sustainable and developed. It highlights the notion of knowledge, the social nature of its construction, the wide range of players involved and the variety of components that have to co-evolve together. It also points to the limited scope of responsibility and ownership of the various players and the dynamic aspects, including positive or negative feedback loops between various chains of activities.

— The (« HOW ») highlights the diversity and plurality of possible approaches to both exploring and narrowing the field of alternatives. These can be large scale, incremental activities with long-term governance. The critical aspect of cooperation and win-win situations are also identified.

### 1.2. *Capturing Knowledge through Sustainability Patterns*

The constitution and structuring of a knowledge base in a specific domain (in this case, fairness) is a long-term and iterative process. One methodical approach is that of **design patterns**, which has its roots in the field of architecture and is used successfully for software systems [Gam+95]. It consists of relating common ways of solving problems with certain characteristics by identifying and analysing practices collected in the field so that they can be reused in a reasoned way.

— **Summary** - An overview of the intent, key dimensions and values.
— **Applicability** - Context for which the pattern is appropriate or not.
— **Content** - Aspects to consider for a requirement derived from this pattern, through clear attributes, capabilities, characteristics or qualities of the system.
— **Archetype** - A pattern description expressed in a generic way.
— **Examples** - Typical instantiation, based on case studies which were used to inspire it.
— **Discussion** - More detailed explanations and assistance with implementation. This part will not be systematically developed in this article. due to limited space.
— **Related patterns** - Possible interactions/combinations with other patterns.

Note that these patterns can be criticised for not offering a clear separation between the sections describing the problem and the solution. The problem is covered by the *Summary* section (in particular the intention) and *Applicability* (context), while the rest of the pattern focuses on the solution.

### 1.3. *Structuring and Reasoning about Knowledge using a Sustainability Meta-Model*

The pattern mechanism has some structuring possibilities, in particular via classifications and links between related patterns. However, these are limited by the textual approach to description, particularly at the archetype level. The latter requires reference to important concepts identified in the foundations, such as goals, dimensions, values and activities.

In order to capture these notions precisely, a meta-model has been developed [PF12] [PF13]. This describes the different dimensions of sustainability, values, activities, regulations and metrics, as illustrated in Figure 1. The left-hand side of the figure describes the semantic structure of the concepts and their associations using a UML class diagram, while the right-hand side gives the graphical syntax. The *Goal* (or objective) is a high-level property of the socio-technical system under consideration. The *Dimension* is a specific point of view of sustainability which may be social, economic, technical or personal in nature. It is broken down into a set of *Values* which carry meaning for the dimension concerned. This value is evaluated by indicators and can be broken down hierarchically into sub-values. An indicator is a qualitative or quantitative metric linked to a value. A *Regulation* is an external element which affects a value. An *Activity* is an operational means of achieving or influencing a value.

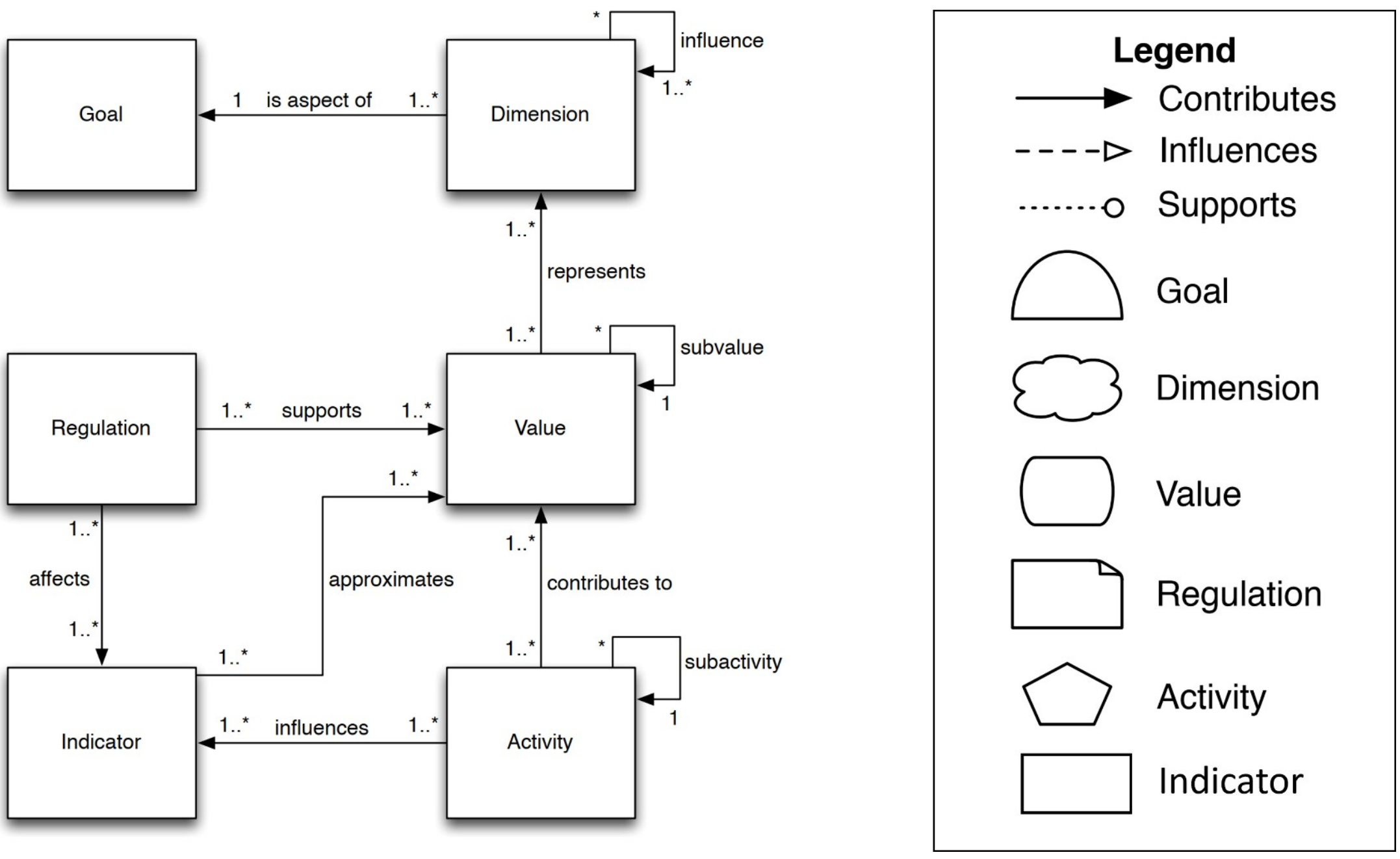

**Figure 1:** Sustainability meta-model : semantics (left) and graphical syntax (right)

This frame of reference has been used in several case studies, one of which concerns fairness [HC15] and is shown in Table 1. Figure 2 illustrates the decomposition of a socio-cultural fairness value resulting from this work. This type of fragment feeds into the pattern identification process.

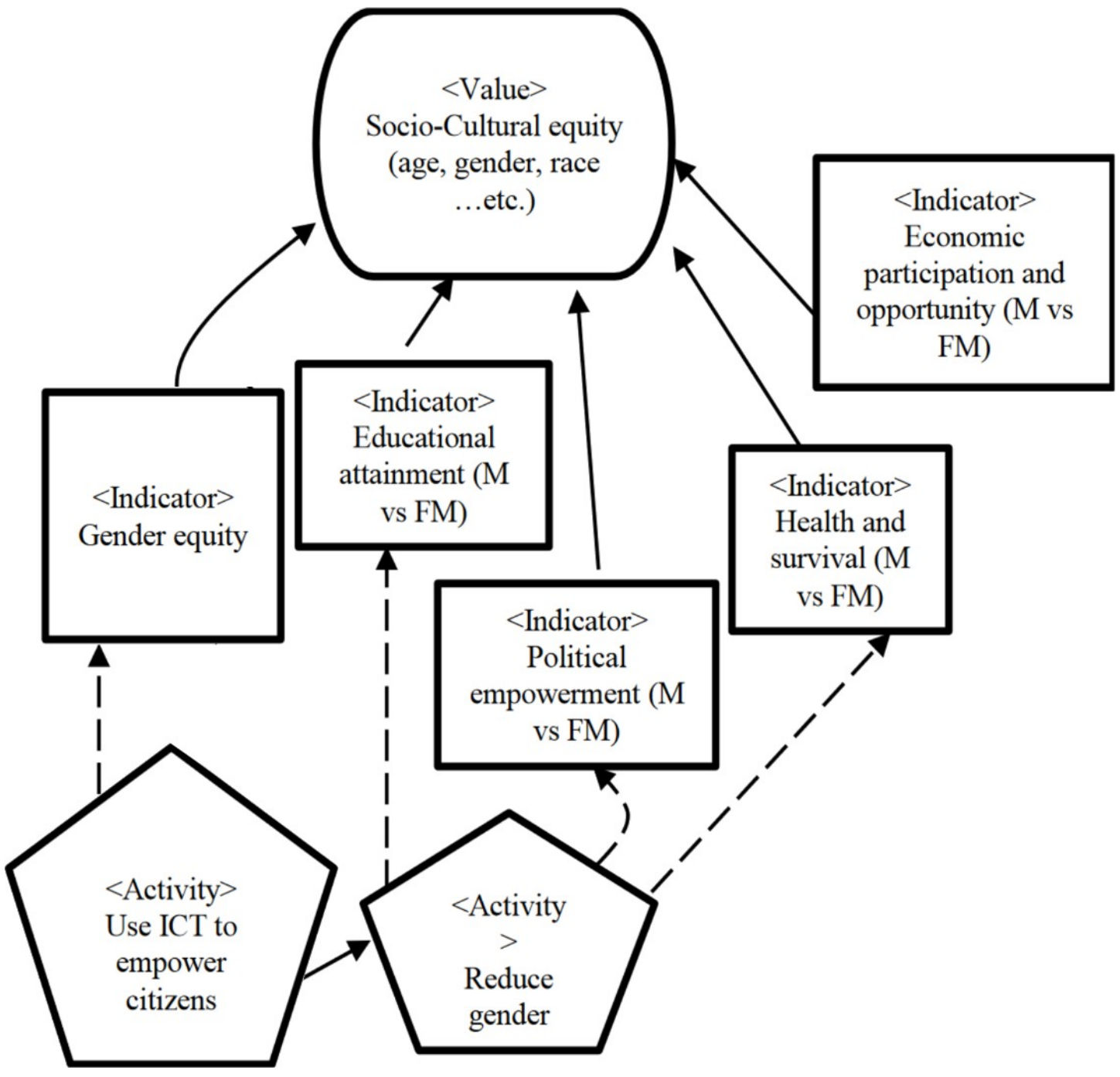

**Figure 2:** Fragment relating to a socio-cultural fairness value [HC15]

This approach contributes to understand the complexity of systems through a combination of heterogeneous models integrating various economic, social and environmental dimensions. It aims at helping scientists and decision-makers to work together [Kie+20]. These allow more specific modelling of sustainability than generic Requirements Engineering (RE) notations such as i* [YM97], KAOS [Lam09] or URN/GRL [ITU12]. However, the latter can also be used, even if they are more general : some of them will be included in the inventory described below. Techniques for building such RE models can also be implemented, in particular to ensure their completeness or robustness [Lam09]. Some of these will be detailed in our methodology presented in Section 2.

## 1.4. *Inventory of Fairness-related Case Studies*

The process of populating and structuring the knowledge base is gradual and must be fed by the analysis of existing systems. To this end, we have listed and analysed numerous case studies with a fairness dimension that have been published in the literature. Some of these are directly related to our experience in eliciting and modelling fairness requirements. All these cases are characterised in Table 1 in terms of domains, topics addressed, sustainability dimensions involved, notations implemented, one or more references and interesting aspects identified.

| Id | Domain | Subject | Dimension(s) | Notations | References | Questions |
|---|---|---|---|---|---|---|
| 1 | Housing | Physical | social economic | KAOS | [PD17] | non-discrimination, new/old building, information availability |
| 2 | Housing | Logement social | social economic | metamodel | [Jon20] | distributive justice, affordability |
| 3 | Food | Fishing quota | environmental social economic | textual | [Doe+16] | computation of equitable quota |
| 4 | Health | Day hospital (oncologie) | social economy | KAOS | [PL18] | medical impact, ethics, staff load management |
| 5 | Health | Night shifts | social economic | textual | [PL18] | juridic rules, people profiles, transparency |
| 6 | Health | Early childhood care | social | KAOS, BPMN | [Chi+15] | links between operation and stratÃl'gy, evolution, enterprise architecture |
| 7 | Services | Accessibility | social | metrics | [PVS20] | non-discrimination, physical/virtual visit, COVID |
| 8 | Services | eCommerce | social economic | GRL | [PT07] | trust, honesty |
| 9 | Services | Decision making | economy | conceptual dynamic | [MG11] | impact of fairness on consumer choice |
| 10 | Services (public) | Water management (Australia) | environmental social economic | textual | [SN06] | distributive justice |
| 11 | Society | Genry equality | social | textual | [Bur08] [San12] [DL13] | WFTO model, monitoring through capability model |
| 12 | Society | Equality | social | metamodel | [HC15] | gender, socio-cultural, social equality |
| 13 | Society | Tax | social economic | textual | [Saa10] | juridic, understanding |
| 14 | Transport Logistics | Procurement | economy social | conceptual | [Fea+12] | fairness, governance, collaboration and access to resources |
| 15 | Transport Logistics | Car pooling | social, environ. economic | KAOS, BMC | [Chi+15] [PLG18] | rewarding customer sharing, driver load, legal rules |
| 16 | Transport/ Logistics | Shared cars/bikes | social, environ. economic | conceptual | [Ma+18] | co-evolution, innovation contribution |

**Table 1:** Use case caracterisation

Analysis of the table shows an interesting coverage of a series of areas of human activity such as housing, services, transport and logistics, health and society in general. A standard classification has not been used here, as this raises the question of the reference framework to be considered and possible biases (for example, reference frameworks for purely economic activities). The various dimensions of sustainability are generally covered in combination, with a predominance of social aspects, which is not surprising as it is directly linked to fairness. It should be noted that a more detailed analysis could attempt to verify the presence of different people profiles, for example to ensure coverage of age and gender categories. A fairly wide range of notations had to be analysed : some were close to the proposed reference framework (same meta-model or goal-oriented notations) while others were based on more specific conceptual notations or even purely textual notations which required a little more analysis work. The various issues identified in the last column have led to the discovery and documentation of specific patterns described in the rest of this article.

## 2. Proposed Methodology

The followed approach is to set up a knowledge base adapted to the analysis of the fairness of socio-technical systems is based on a process consisting of the following four stages, which are detailed in the remainder of this section.

1. As a preliminary step, we have defined a modelling repository based largely on the state of the art presented in the previous section, with the addition of extensions enabling richer and more modular modelling. A more targeted analysis of the IS component is also highlighted.
2. The next step is to carry out pattern discovery work based on the analysis of the various case studies, their modelling and the identification of recurring patterns in order to capture them in a reusable way using our repository. To illustrate this approach, a more specific case study will be analysed : the fairness of clinical itineraries in oncology [PL18].
3. In order to make it possible to use a growing library, it is also necessary to propose an appropriate structure. To this end, a standard change management process is proposed.
4. Finally, the user must understand how to exploit the patterns available in the library in order to use them effectively when analysing a new socio-technical system and its IS component.

It should be noted that this approach can be generalised to a wider range of properties related to sustainability, but in the context of this work, we have focused on the construction of a knowledge base restricted to the property of fairness.

### 2.1. *Extensions of the Sustainability Meta-Model and Pattern Template*

**At the level of the meta-model of sustainability notations**, the purpose of the proposed meta-model is to target sustainability with, in our case, a more limited application to fairness properties. It is of course possible to use more generic RE notations such as i* [YM97], KAOS [Lam09] or URN/GRL [ITU12]. However, the intention was to align with existing work specific to sustainability such as [PF13]. Conceptually, the meta-models are largely compatible with a refinement approach from properties (here oriented towards sustainability) towards an operationalisation in the form of activities with an intermediate concept of value and a classification by sustainability dimension. This has made it possible to integrate case studies carried out using more generic EI notations. Conversely, some EI analysis techniques can also be transposed to our analysis approach. In fact, we were faced with several limitations that we were able to resolve by means of conservative extensions to the initial meta-model. These are shown in grey in Figure 3, which also uses French terminology, unlike the initial meta-model shown in Figure 1.

**A first limitation addressed is the lack of a mechanism for initiating interesting reasoning about the properties of a system in a non-ideal environment** and thereby enriching the model. To remedy this, we have incorporated, in Figure 3, the concept of *Obstacle* by defining it as obstructing a *But* and,

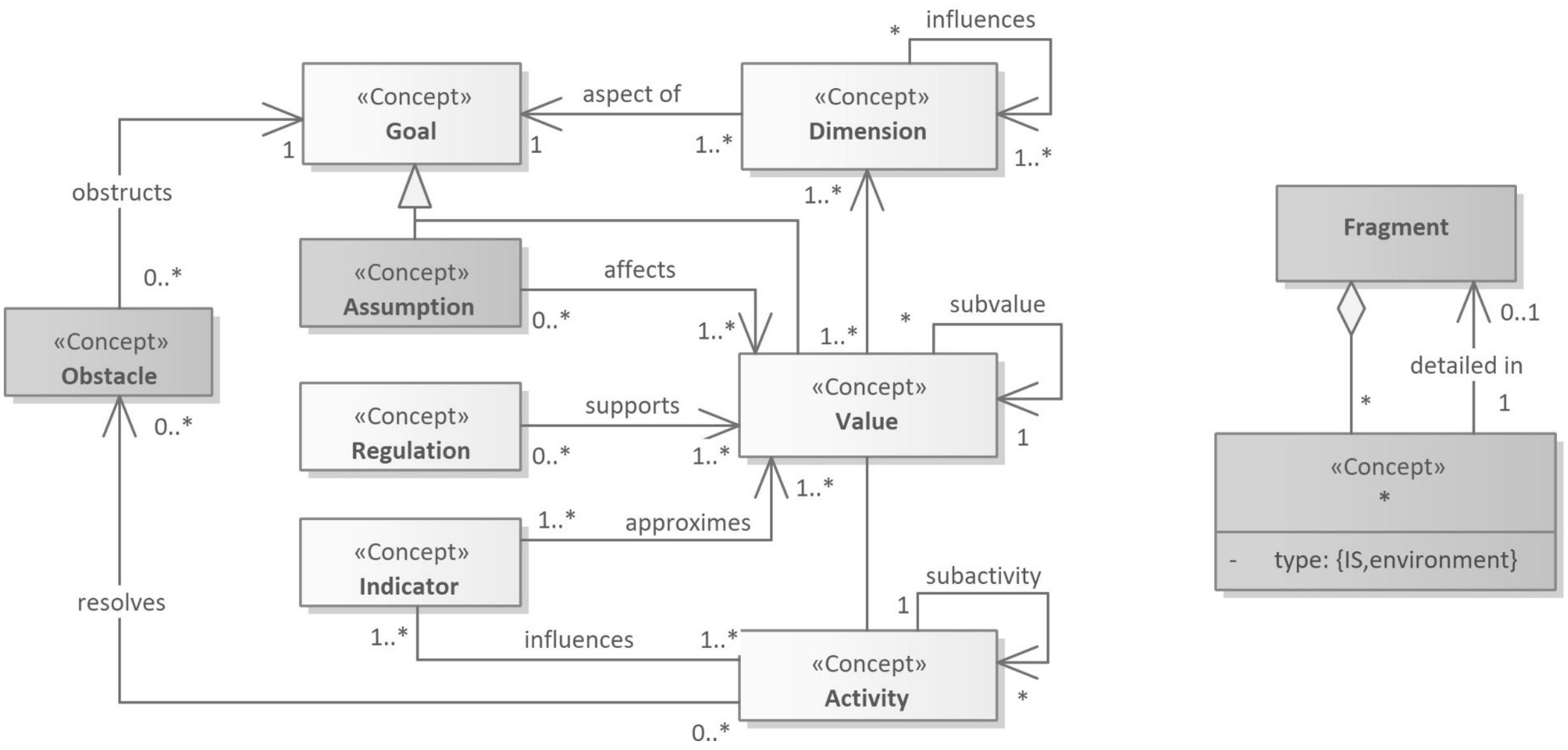

**Figure 3:** Meta-model extended with obstacle, assumption and fragment (grey concepts)

by inheritance, also a *Value* which is interpreted as being of the same nature as a *But* but specific to a *Dimension*, as well as the new concept of *Assumption*. This concept is introduced to capture a behaviour which must be satisfied in order for the *Value* concerned to be guaranteed, without the system imposing it. It makes it possible to clarify a realised idealisation and possibly to question it later. In order to resolve an *Obstacle*, a specific *Activity* can be introduced.

We note here the influence of RE on these extensions, enabling better reasoning on the realism of models. In order to discover obstacles and propose resolution strategies, techniques developed in RE are directly applicable. Strategies for identifying obstacles include fairly systematic but also very formal techniques for regressing the negation of a goal through the properties of the domain in order to identify causes. Other strategies are based on fairly generic obstacle patterns ; for example, a famine pattern is highly relevant for generating violations of fairness of access to a resource. Finally, less formal strategies are based on heuristics and techniques similar to risk or fault tree analyses. For example, we can establish typologies of causes that may lead to access to a resource : its failure, its use by another agent, a delay in planned availability, confusion with another resource, etc. In terms of obstacle resolution strategies, various techniques directly inspired by risk analysis are also available : acceptance, avoidance, reduction (which can be achieved by restoring the goal or accepting a weaker goal) or pure and simple acceptance.

**The second major limitation is the absence of a modularity mechanism for breaking down a complex system into smaller fragments**. This immediately limits the complexity of analysable systems. On the other hand, such a notion is also needed to capture interesting fragments and generalise them into patterns that can be characterised and classified for reuse. This limitation has been addressed by introducing the concept of *Fragment* » as shown on the right of Figure 3. A fragment can contain any type of model element detailed in a dedicated diagram. This fragment can then be linked to a model element in a diagram that imports it. It can also be made generic, forming a reusable *Patron* that can be instantiated in a specific context.

**Finally, a third limitation is the lack of distinction between the IS and its environment in the socio-technical system**. One of the objectives is to determine how the IS can support fairness properties. A specific attribute can be associated with any concept in order to characterise its impact on the IS or its

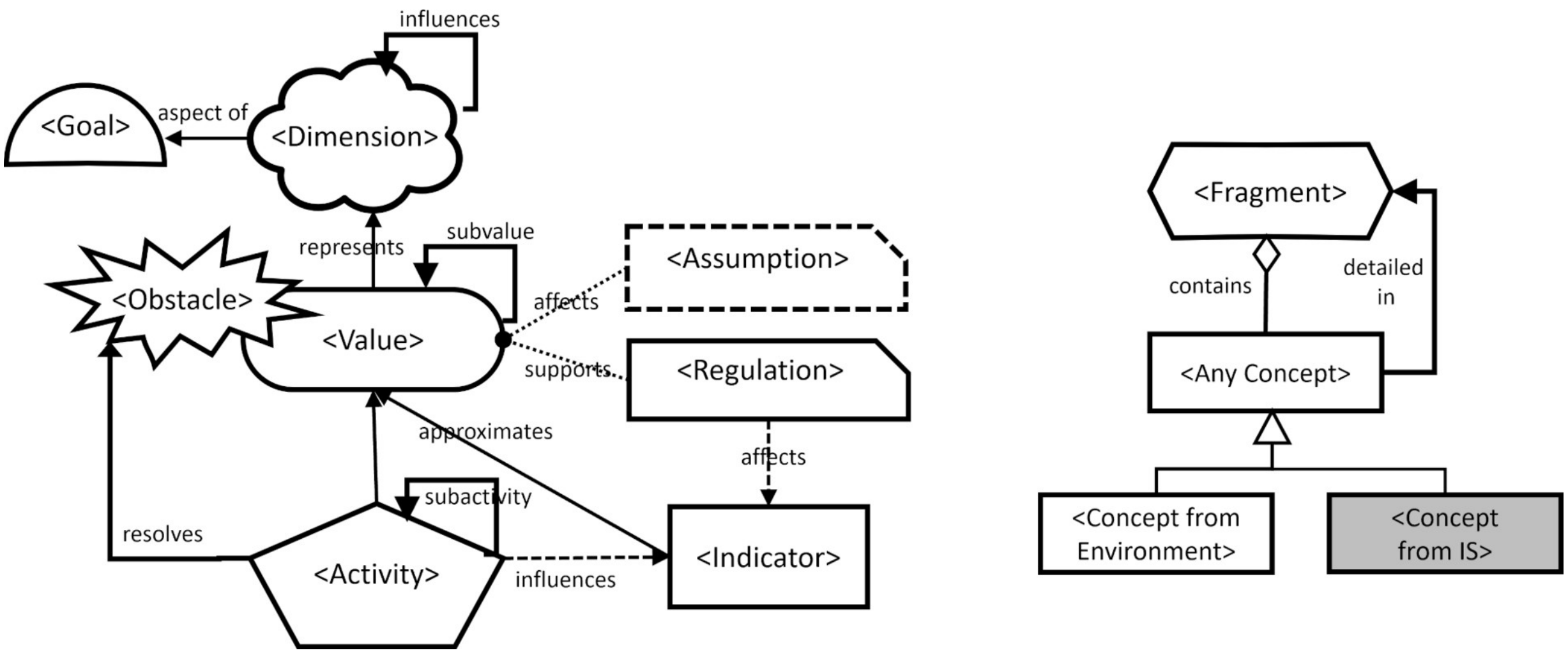

**Figure 4:** Extended graphical notations

environment, or even the « machine »=and the « world », to use the terminology of the problem frames of [Jac01]. In order to make this attribute explicit visually, a grey background is associated with all concepts relating to the IS, as illustrated in Figure 4. This figure also proposes graphical notations for our additional concepts, in particular an obstacle is represented graphically by a small explosion which can be positioned directly on the obstructed concept (here a value).

Further extensions were possible but have not been implemented at this stage, for example the notion of alternative. These will be discussed in Section 5. **At the level of the pattern model**, we have largely remained aligned with the existing pattern proposed by [RR13], enriching it however with the following two elements allowing a better classification of the patterns :

— **Category -** Name of the main (and possibly secondary) category of the pattern according to the typology presented in Section 3.

— **Dimensions -** One or more dimensions of sustainability addressed by this pattern. The proposed classification includes the following dimensions : environmental, economic, social, personal and technical [Pen+14].

## 2.2. *Process for Identifying Fairness Patterns*

The approach is based on the list of case studies presented in Section 1.4. Each case is analysed using the following systematic approach :

1. Identification of values, regulations or assumptions relating to fairness, including analysis of the concepts of equilibrium, distribution or arbitrage.
2. Analysis of existing models (possibly in related RE notations) or a posteriori modelling depending on the level of information available.
3. Analysis of business indicators in order to determine whether they have an explicit or implicit link with notions of fairness.
4. Identification of recurring strategies between several case studies, for example, in relation to the definition of rules, their transparency, the adherence process, etc.

By way of example, Figure 5 illustrates the modelling approach using these notations on a case study of a care pathway taken from Table 1 in [PL18]. The human, social and economic dimensions are present. On the human side, maximum quality of care is expected. However, due to limited capacity (which cannot be resolved economically), a social value of equitable quality of care for all patients is defined. This means that all patients are treated equitably in order to guarantee the effectiveness of their treatments, which may require compliance with a well-defined timing, expressed through regulation (for example, to avoid increasing the risk in the event of a cancer relapse). Patients are also required to keep their appointments for their own benefit and because rescheduling may have an impact on other patients or increase the burden on staff. Overall, the load is spread fairly across staff with particular attention to overtime.

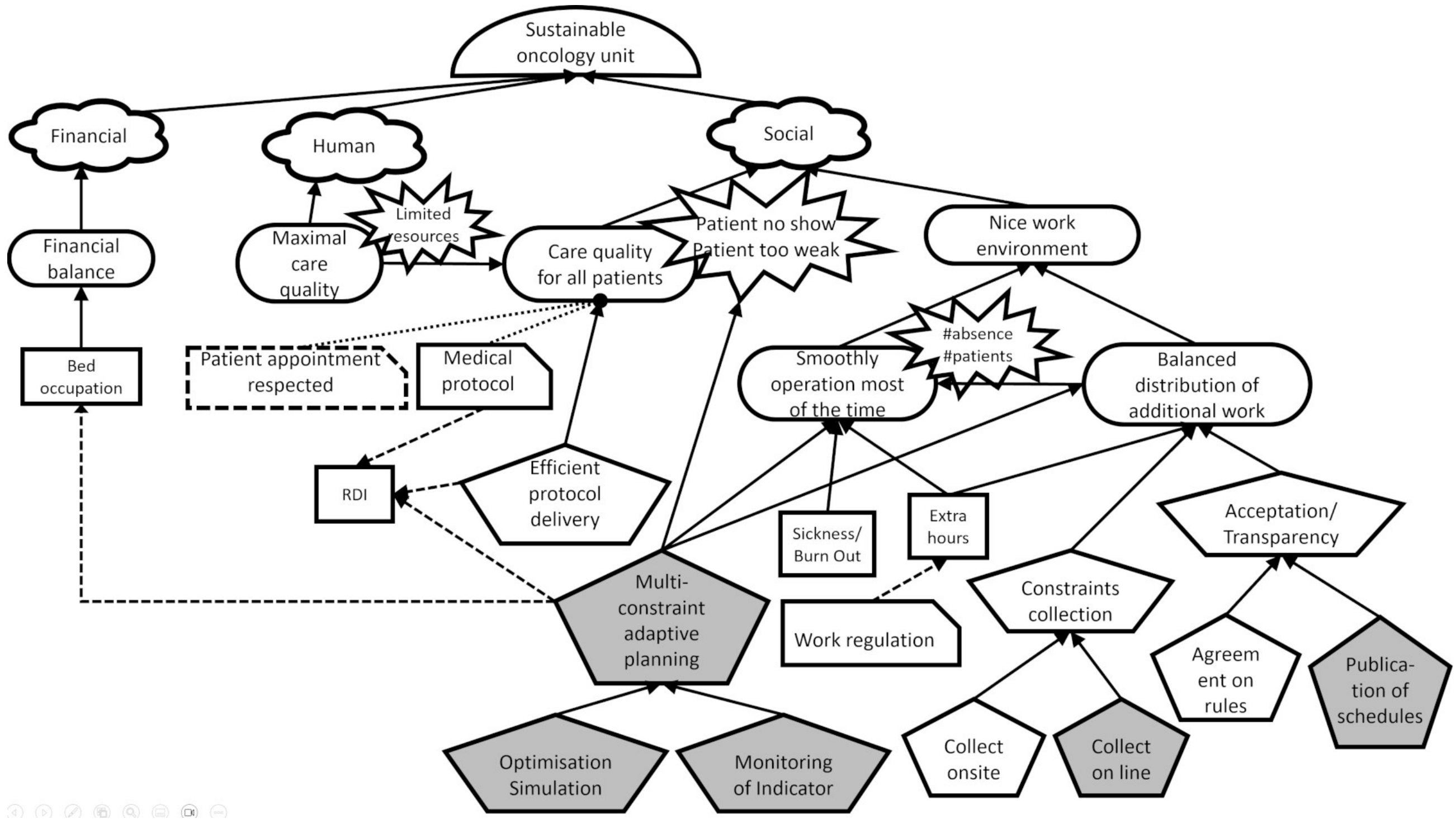

**Figure 5:** Model of a oncology care unit in a day hospital

Two main indicators are used to monitor fairness. The first is the RDI (Relative Dose Intensity). This is a medical indicator that shows the extent to which a patient's pathway adheres to the theoretical pattern in terms of compliance with the required times and doses. It should be as close as possible to 1.0, with minimal variance between patients to ensure fairness. The second concerns the staff workload, for which the distribution of overtime must be monitored. The volume of overtime should be similar for all staff, taking into account their occupancy rate. Analysis of this example reveals various aspects of fairness linked to compliance and adherence. It also highlights traceability based on a series of measurable indicators, some of which can even be published for transparency purposes, for example, for fair planning of overtime hours [PL18].

Analysis of this model has already revealed a series of building blocks that can potentially be generalised and documented in the form of patterns. In fact, elements such as the use of load and simulation indicators are integrated into the violation anticipation pattern, which is also an obstacle avoidance strategy. Elements relating to agreement on rules and publication of schedules feed into the acceptance and transparency patterns respectively. These various patterns were consolidated on the basis of an analysis covering the other cases (according to the process described above) and detailed in Section 3.

## 2.3. *Structuring a Catalogue of Fairness Patterns*

For a pattern catalogue to be usable, it is important that it is not just a list of fragments, but that a relevant classification is proposed in order to facilitate the identification of a specific pattern. For example, for the object-oriented « Design Patterns » patterns are grouped by the function of creating, structuring or managing the behaviour of objects [Gam+95]. As a structure, our case studies have led us to consider an evolutionary process, as it is ultimately a question of managing change in order to achieve transition and/or improve the sustainability of systems, as pointed out in our state of the art [Sav+19]. This organisation makes it possible to naturally follow a continuous improvement process according to the PDCA (or PDSA) cycle « Plan-Do-Check/Study-Act » attributed to Deming [MN10]. It has been applied to ensure the properties of the decision component in the design of DIS (Decision Information Systems) [Ann+06]. The aim is to include recommendations for increasing maturity in terms of fairness and sustainability. Unfortunately, these are not very present in maturity repositories such as CMMI (Capability Maturity Model Integration).

The content of the steps described in Figure 6 is as follows :

— **The design stage** ails at identifying the system's operating rules are drawn up. This involves different types of actors representing the various stakeholders : sponsors, designers, user representatives, etc. In order to guarantee representativeness and diversity, democratic and participative design patterns can be used.

— **The adoption stage** focuses on ensuring that the designed system is accepted by the users in all their diversity. The stakeholders involved here are the users of the system. Participatory design already contributes to this upstream, as do acceptance and transparency initiatives.

— **The implementation stage** is more operational and technical. It is, moreover, the one that will involve IS the most. The players involved are the operationally-oriented managers and the administrators, including with regard to the IS. There will be measures to guarantee access to specific targets, measures to anticipate breaches of fairness and contingency measures.

— **The evolution stage** is used to analyse the extent to which the system is deviating from the expected results for various reasons (ineffective measures, changes in the context, etc.) and to initiate a new cycle. The players involved are analysts and more strategic managers who rely on metrics derived from the various patterns. Specific patterns concern the objective analysis of data to avoid bias, particularly in machine learning processes. A co-evolution approach can also be implemented in conjunction with governance.

— **The governance cross-cutting activity** is responsible for overall management. The players involved are the managers responsible for the various systems in operation. Fairness can be ensured by following the main principles of distributive justice or substantial freedom mentioned in the introduction. Diversity in governance is also a key element.

## 2.4. *How to use the Fairness Pattern Catalogue*

We describe here a fairly broad scenario of appropriation. Certain steps may be unnecessary for users who are already familiar with the system, or optional for novice users.

1. A recommended first step before using the pattern library is to be familiar with the meta-model and its graphical notations and to understand a modelling example. This material is presented in

Section 2.

2. Next, before any analysis, it is advisable to understand the structure of the library, also described above, and to go through a few patterns to understand how they are organised and how to identify them via their position in the structure and the dimensions supported or the description of their intention.
3. An initial identification can then be confirmed by examining the description of the archetype. If the pattern is unsuitable, an alternative pattern from the discussion or the associated patterns can be consulted.
4. In the modelling phase, the analyst follows his or her own approach, starting with the identification of key dimensions and associated values. The analysis carried out may already lead to the consideration of certain patterns by searching the library. This stage can also take place at the most detailed level of a specific activity.
5. The inclusion of a pattern leads to the model being enriched with certain elements proposed in the archetype, as not all the elements are necessarily relevant in the context under consideration or may have been taken up in another form. If necessary, the model can be further restructured.
6. The process is continued, in particular by refining the activities or considering obstacles. This may also lead to the identification and application of patterns.
7. A regular and experienced user may be led to identify undocumented recurring elements in the library and augment it for personal or organisational purposes, or even by sharing them more widely, thus contributing to the enrichment of the library.

## 3. Catalogue of Fairness Patterns

This section documents the patterns discovered from the cases described in Table 1. Those currently identified are presented according to the model described in section 1 with the additions presented in Section 2. The archetype is described using the augmented notations introduced in section 2. The patterns are organised in the form of a **catalogue** whose overall structure is illustrated in Figure 6. It is made up of 4 key stages forming a virtuous circle enabling a continuous improvement process presented in Section 2.3. This cyclical process is completed by an activity at the centre of the cycle, which manages its governance..

The figure positions our various patterns (hexagons) in an area indicative of their contribution. This may correspond to a specific activity, but it may also form a link between several activities. For example, the violation anticipation pattern fits into an implementation stage framework but also goes beyond this, by enabling a form of self-adaptation and fuelling the need for evolution, as we will see in the details of this pattern.

The rest of this section describes a selection of our patterns. We review them in a non-exhaustive way, introducing those that are used as part of our validations, in particular, those that more specifically involve information systems. In this respect, it should be remembered that the activities under the responsibility of the IS are represented with a grey background.

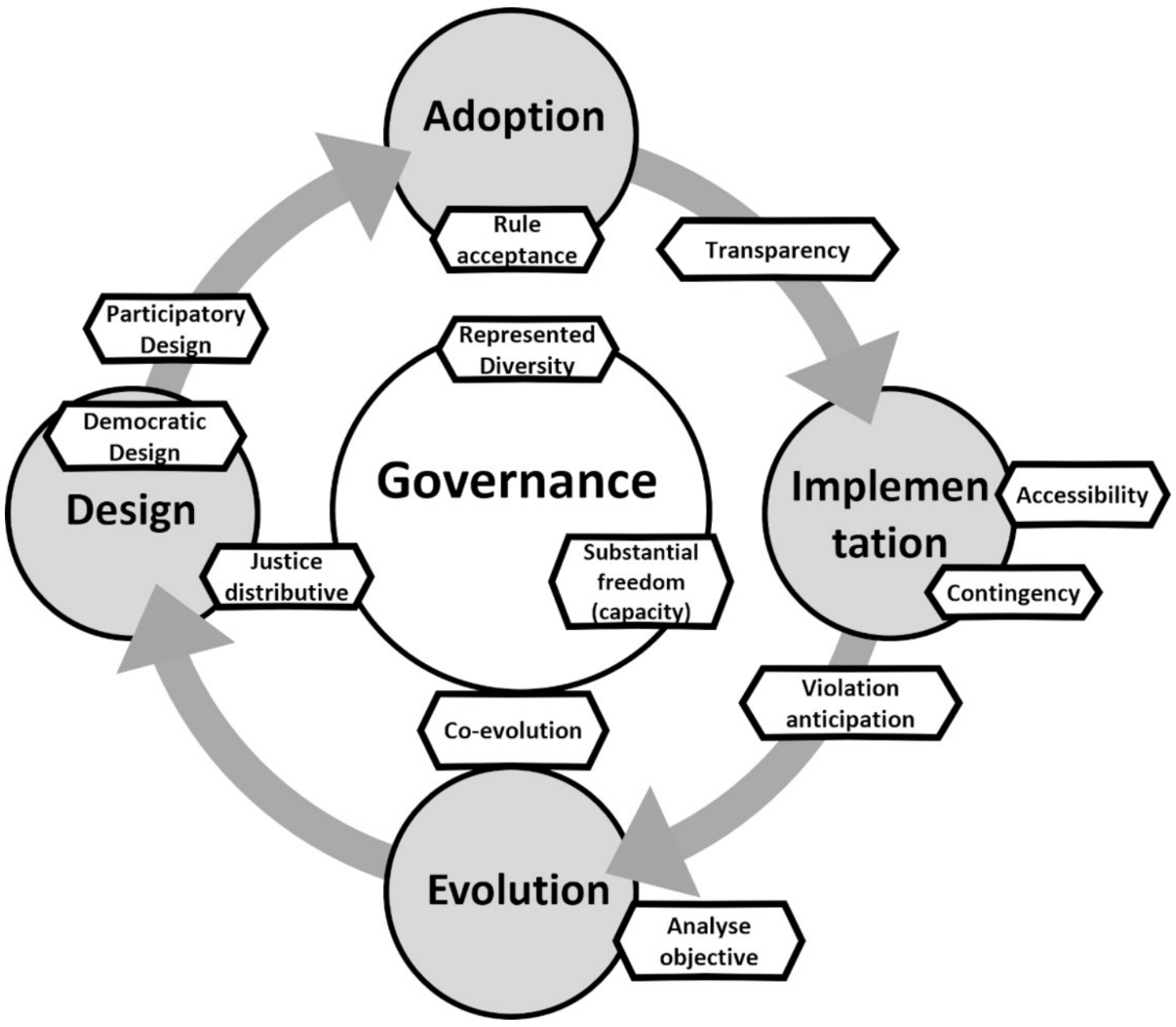

**Figure 6:** Structure of our Fairness Pattern Library

### 3.1. *Distributive justice pattern*

**Category** - governance (design).

**Dimensions** - Ãl'conomic and social.The pattern supports different value models, the applicability of which depends on the context and involves different indicators. **Summary** - Distributive justice concerns the fair distribution of resources (material or services) between the various members of a community. It takes into account the total quantity of goods to be distributed, the distribution procedure, particularly in relation to equality, equity, need and responsibility, and the resulting distribution model. **Applicability** - The pattern supports different value models, the applicability of which depends on the context and involves different indicators.

**Content** - The characteristics and needs of the community must be assessed in order to select the values to be distributed.

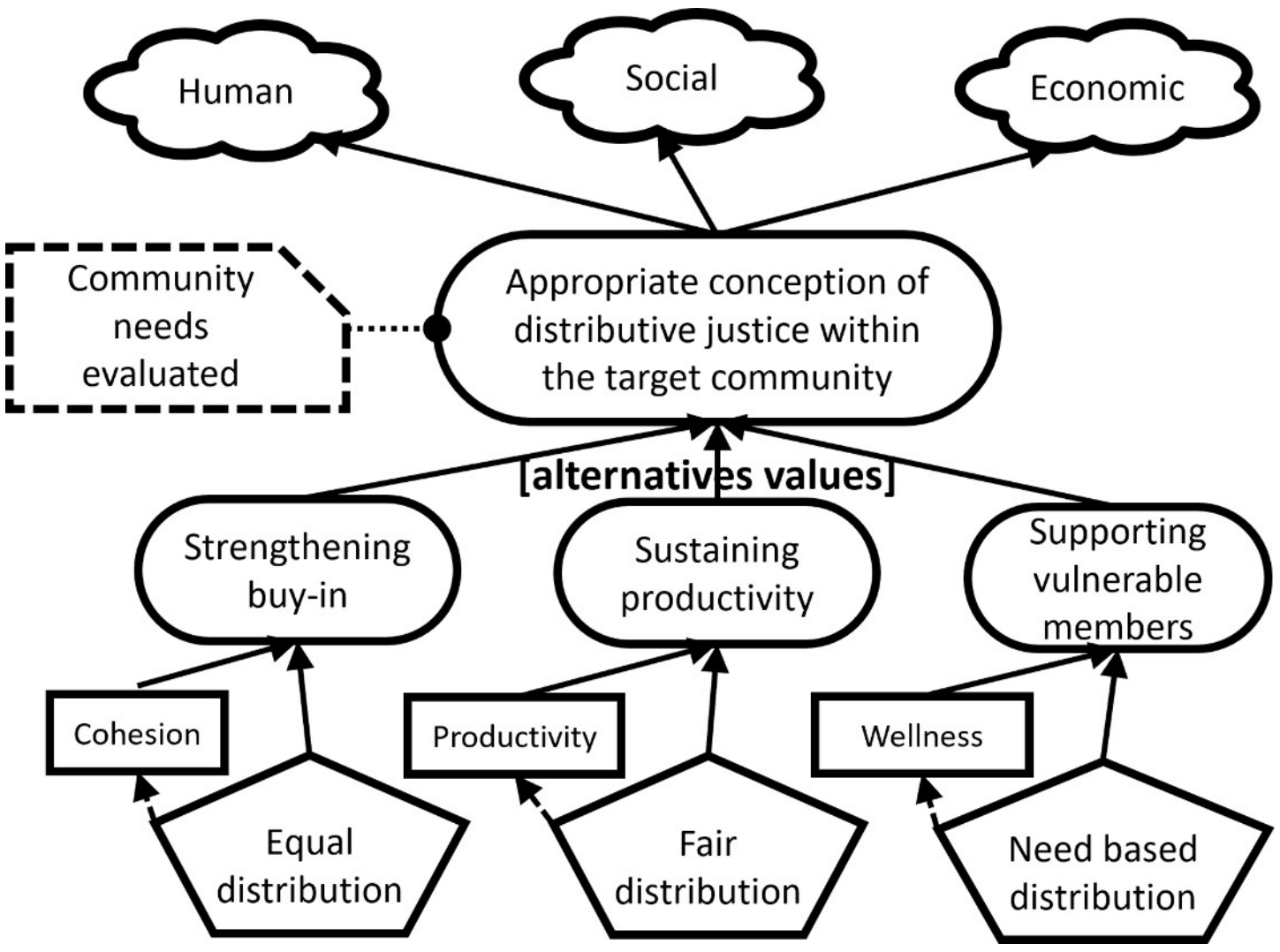

**Figure 7:** Distributive justice pattern

Archetype - Figure 7 shows the breakdown of value models with dedicated activities and indicators. Equal distribution emphasises a sense of belonging. Fairness promotes motivation to produce through proportional remuneration. Needs-based distribution ensures that common basic and essential needs are met. It contributes both to individual well-being and to a reduction in the risk of criminal and political violence [Mai13].

**Examples** - Labour regulations, tax rules, fishing quotas.

**Related patterns** - Adoption of fair, participative rules (not detailed), substantial freedom (in addition).

## 3.2. *Substantial freedom pattern (or capability pattern)*

**Category** - governance (implementation).

**Dimensions** - Individual = this pattern targets a better personal accomplishment.

**Summary** - This pattern proposes strategies to respond to the limits of distributive justice, which is based on the equality of primary social goods but is not sufficient to guarantee that individuals will enjoy the same freedom of action. It puts forward the notion of capability, which corresponds to the effective possibility of carrying out certain rewarding acts, taking into account personal characteristics and external factors.

**Applicability** - The pattern supports different value models, the applicability of which depends on the context of the people involved. It is based on specific indicators such as the HDI (Human Development Index) and gender equality indicators such as the GEM (Gender Empowerment Measure). This contrasts with economic indicators such as distribution keys or quotas in distributive justice approaches.

**Content** - The characteristics and needs of the community must be assessed in order to select the values to be distributed.

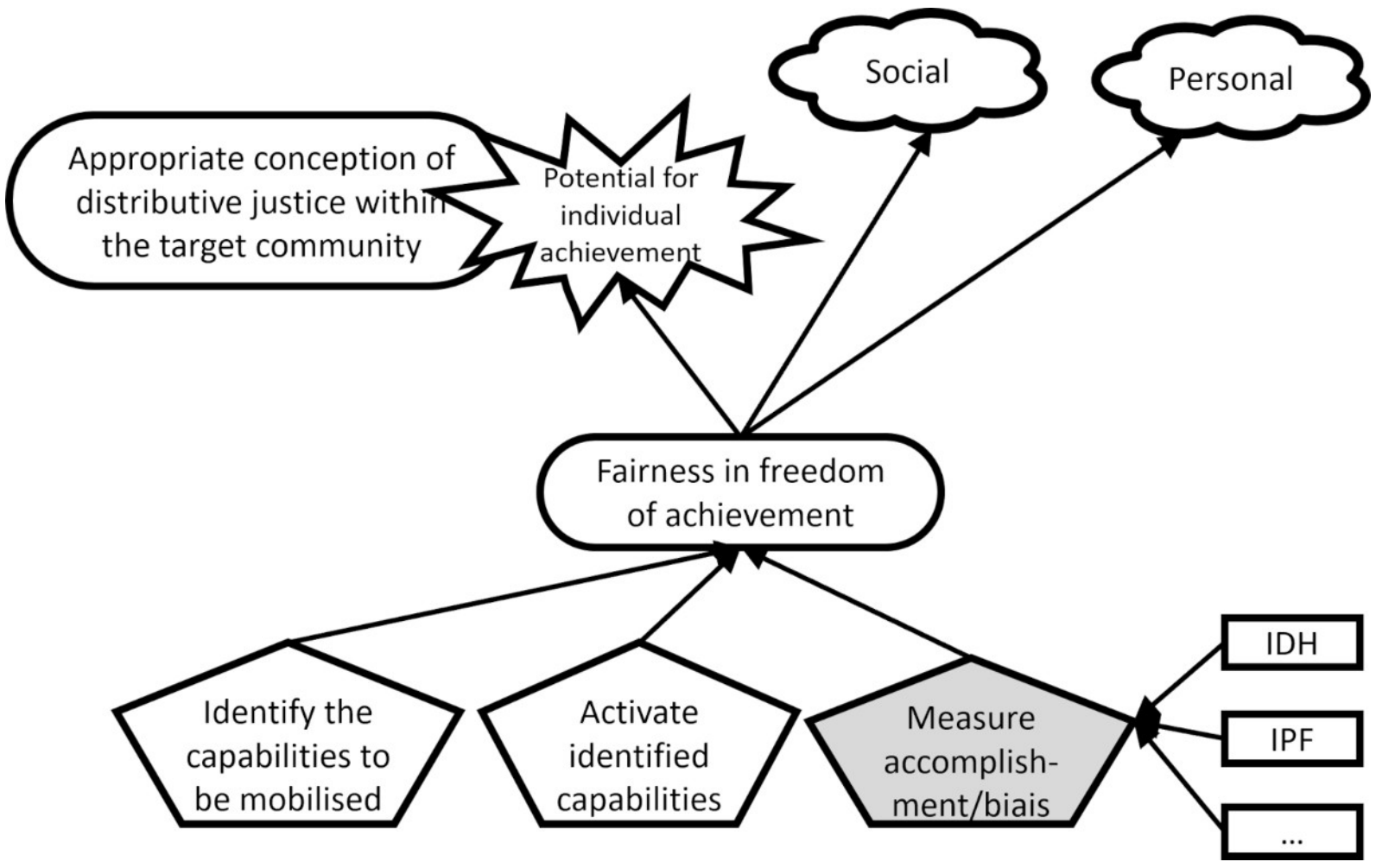

**Figure 8:** Substantial freedom pattern (or capability pattern)

**Archetype** - Figure 8 shows the structure of the pattern, which is a response to the obstacle of personal fulfilment in distributive justice. Its implementation is fundamentally anchored in the virtuous circle in the sense that it involves matching the capabilities to be mobilised with those that can be mobilised by encouraging diversity. The process is essentially based on the HR (Human Resources) aspect. The IS is useful for analysing achievement indicators. Beyond this, part of the strategy also consists of transforming the system and therefore activating patterns such as co-evolution.

**Example** - gender equality (WFTO).

**Related patterns** - Distributive justice, co-evolution.

### 3.3. *Rule acceptance pattern*

**Category** - Adoption.

**Dimensions** - The characteristics and needs of the community must be assessed in order to select the values to be distributed.

**Summary** - This pattern deals with the acceptance of the fairness of the rules applicable to a given system. It involves various dimensions and may rely on mechanisms such as consultation, voting or consensus to achieve collective acceptance.

**Applicability** - Initial or subsequent acceptance phase. For example, after detection of evolution or degradation.

**Content** - The rules need to be tailored to the target audience. Important indicators such as diversity, commitment and the level of precision of the rules must be quantified.

**Archetype** - Figure 9 shows the decomposition of the main acceptance values into three sub-values with their own action. A global time milestone is used to ensure that the rules are understandable, understood and accepted. Iterations are possible to achieve acceptance.

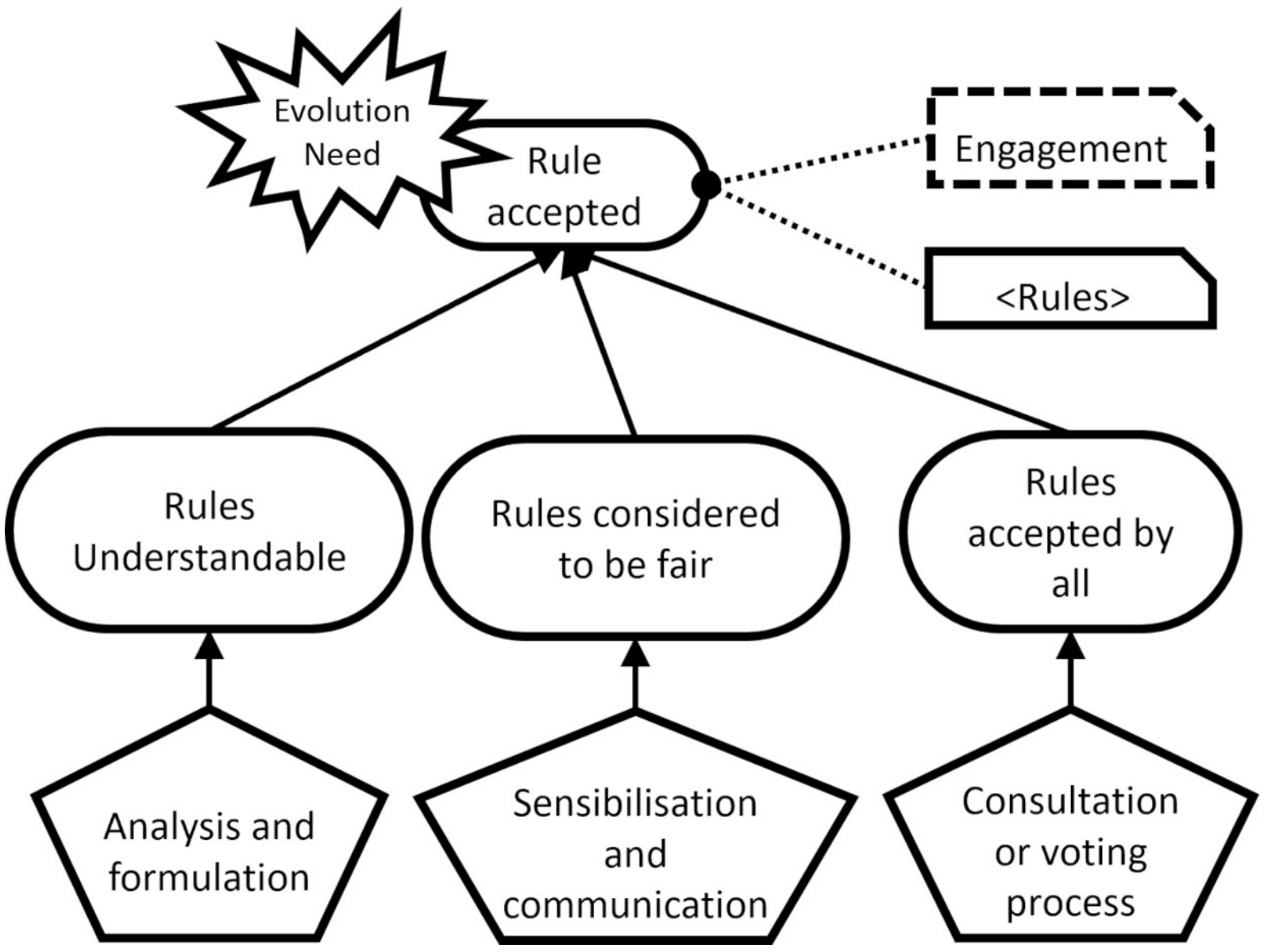

**Figure 9:** Rule acceptance pattern

**Examples** - labour regulations, tax rules.

**Related patterns** - Participatory design (favors adoption).

## 3.4. *Transparency pattern*

**Category** - Adoption (implementation).

**Dimensions** - All dimensions can be considered with a focus that is specific to the field under consideration. For example : social and economic aspects in planning systems, environmental aspects in industries with a potential environmental impact. The technical dimension is also present at the level of implementation within an IS.

**Summary** - This pattern helps ensure transparency of operations regarding certain information that needs to be shared between the target audience, in relation to fairness in our case.

**Applicability** - This pattern ensures the transparency of operations with regard to certain information that needs to be shared between the target audience, in our case in relation to fairness.

**Content** - Indicator to be made available to the community with a guarantee of its representativeness and the process for obtaining it.

**Archetype** - Figure 10 shows how transparency can be achieved at different levels : through a process audit, software inspection using tests or (open) source code, or by publishing evidence of fairness in operational decisions. It strongly involves IS components.

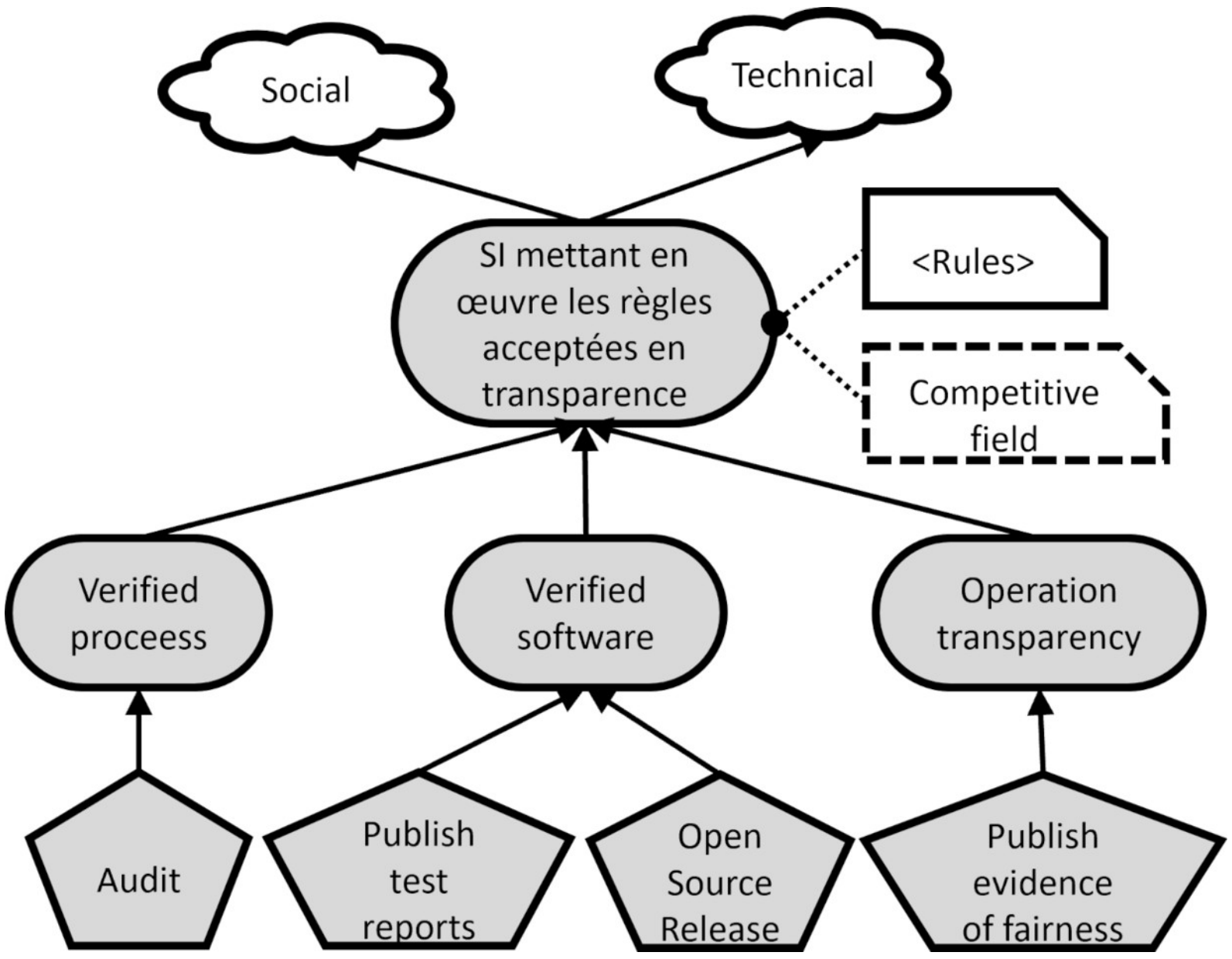

**Figure 10:** Transparency pattern

**Example** - Night shift schedules (including explanations of the rules applied by the software).

## 3.5. *Violation anticipation pattern*

**Category** - Implementation (evolution).

**Dimensions** - All dimensions may be relevant. The indicators to be monitored can be multiple and multi-objective on several dimensions (the most interesting cases).

**Summary** - This pattern deals with the detection of the violation of a condition (in this case fairness, but applicable more widely). When the risk of violation is present, this pattern can anticipate it and thus prevent the violation from occurring thanks to an appropriate measure.

**Applicability** - The property under consideration (in this case fairness) must be predictable using a model (linked to the rules and the domain) and the data to be collected. Implementing corrective measures requires sufficient reaction time. The contribution of IS is both in terms of the effectiveness of deviation detection and the implementation of avoidance measures.

**Content** - Predictive model (oracle, simulator, digital twin, etc.) and data to feed it.

**Archetype** - Figure 11 shows the decomposition of the violation pattern into two main stages : detection and reaction. The detection phase is based on simulation activities and load factor data. The reaction phase can pursue several strategies such as trying to increase resources, balancing or postponing the load to ensure that the capacity to cover all requirements is restored without generating an unfair situation. The IS provides effective support for the monitoring of indicators and the reactive implementation of adjustment activities.

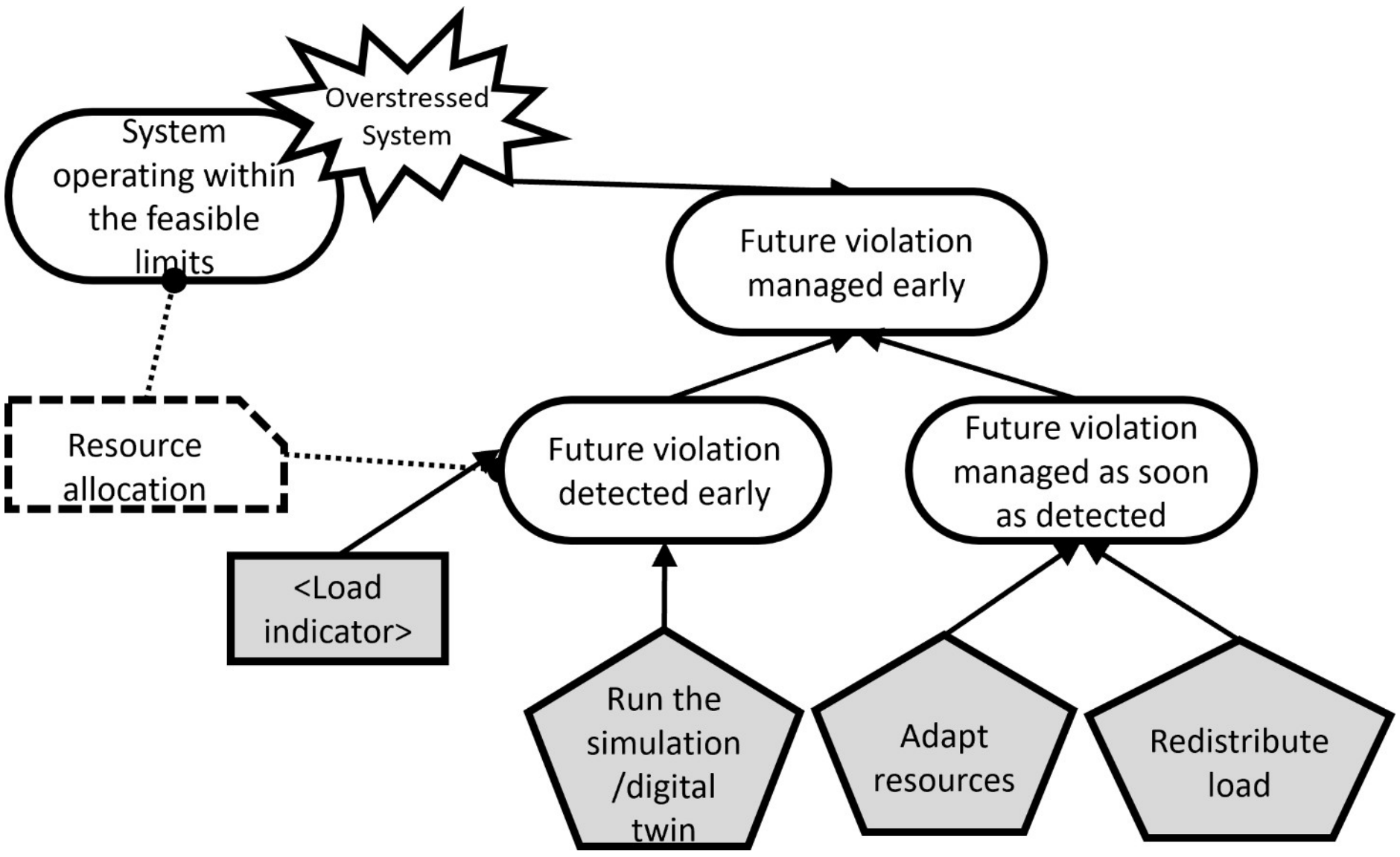

**Figure 11:** Violation anticipation pattern

**Example** - Hospital capacity management, human resources, quotas.

**Related patterns** - Fair rule (indicator to monitor).

## 3.6. *Co-evolution pattern*

**Category** - Evolution (governance).

**Dimensions** - All dimensions may be relevant. At least two dimensions need to be considered for co-evolution.

**Summary** - Co-evolution consists of activating synergies between several areas in order to move the lines simultaneously, thus avoiding the barriers that block strategies anchored in a single area. It is accompanied by innovation activities and an action plan that will fuel new iterations.

**Applicability** - Useful for problems involving multiple interrelated areas.

**Content** - An extended business framework model is useful to support the approach, for example, a circular economy framework integrating environmental and social dimensions [Cir22]. The pentagonal diagram of effect chains proposed by SusAF (Sustainability Awareness Framework) is also highly recommended for identifying synergies [Dub+20].

**Archetype** - Figure 12 illustrates the approach, which is based on an assessment of the situation and its (im)balances, the means of multidisciplinary co-innovation, and public validation, leading to the identification of synergies.

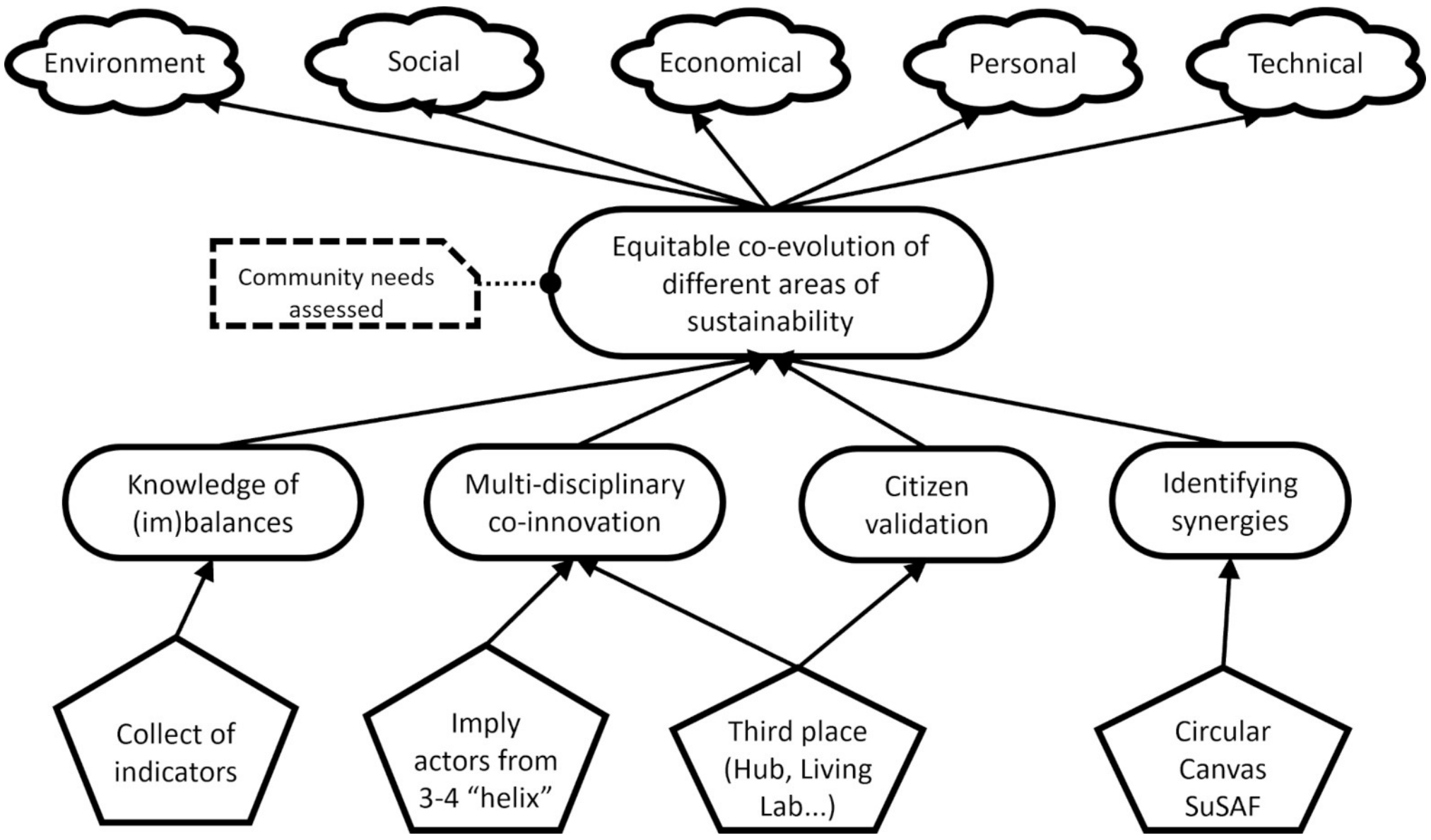

**Figure 12:** Co-evolution pattern

**Example** - Co-evolution of urban mobility (taking into account changes in population, urban planning, public transport, working patterns, etc.).

## 4. Validation

In order to validate our catalogue of patterns, we modelled several case studies using the approach described in section 2.4 to guide the development of a model that incorporates as many fairness mechanisms as possible. We used two case studies :

The first concerns the management of the crisis linked to the COVID-19 pandemic. We analysed it in two successive and distinct phases : the containment phase and the vaccination phase. This case was not included in the list used to draw up the catalogue (see Table 1). In this validation, we are only exploring the overall socio-technical system without going into the details of the underlying IS, for example, to monitor vaccinations. The aim is above all to illustrate the structuring mechanisms and the identification phase, particularly in terms of frequency, context and the type of pattern applied. We will limit ourselves to annotating the places where patterns are interesting to apply, but without unfolding them in the initial diagram, or examining how to make any adaptations : this point will be considered in the second case.

Our second case study concerns a system of medico-social monitoring of early childhood in French-speaking Belgium, which dates back to the First World War [ONE19]. Unlike the other case study, this one was the subject of a prior detailed analysis by us and was published as a representative example in terms of sustainability [Chi+15] but without any specific tool for fairness. The modelling was carried out independently of this initial work, which was based on use case and business process notations. In this validation, the level of detail is pushed further in order to illustrate how the patterns make it possible to identify requirements on the IS underlying certain activities in order to ensure the values linked to fairness. The instantiation and eventual restructuring stage will be explored in greater depth.

The figures that follow are annotated with speech bubbles that are not part of the notations but are more part of the validation process : they point to concepts where a fairness pattern seems interesting to apply. They can be any concept whose formulation or associations might suggest a pattern.

### 4.1. *Pandemic COVID-19 - confinement phase(s)*

Figure 13 shows our global model with values and activities according to the meta-model adopted and ratings from the perspective of fair and sustainable crisis management. It should be noted that this initial model was drawn up in the summer of 2020 at a time when vaccines were not yet available. However, it included a pointer (represented by a hatched texture) to a fragment detailing the vaccination phase, which was analysed at a later stage.

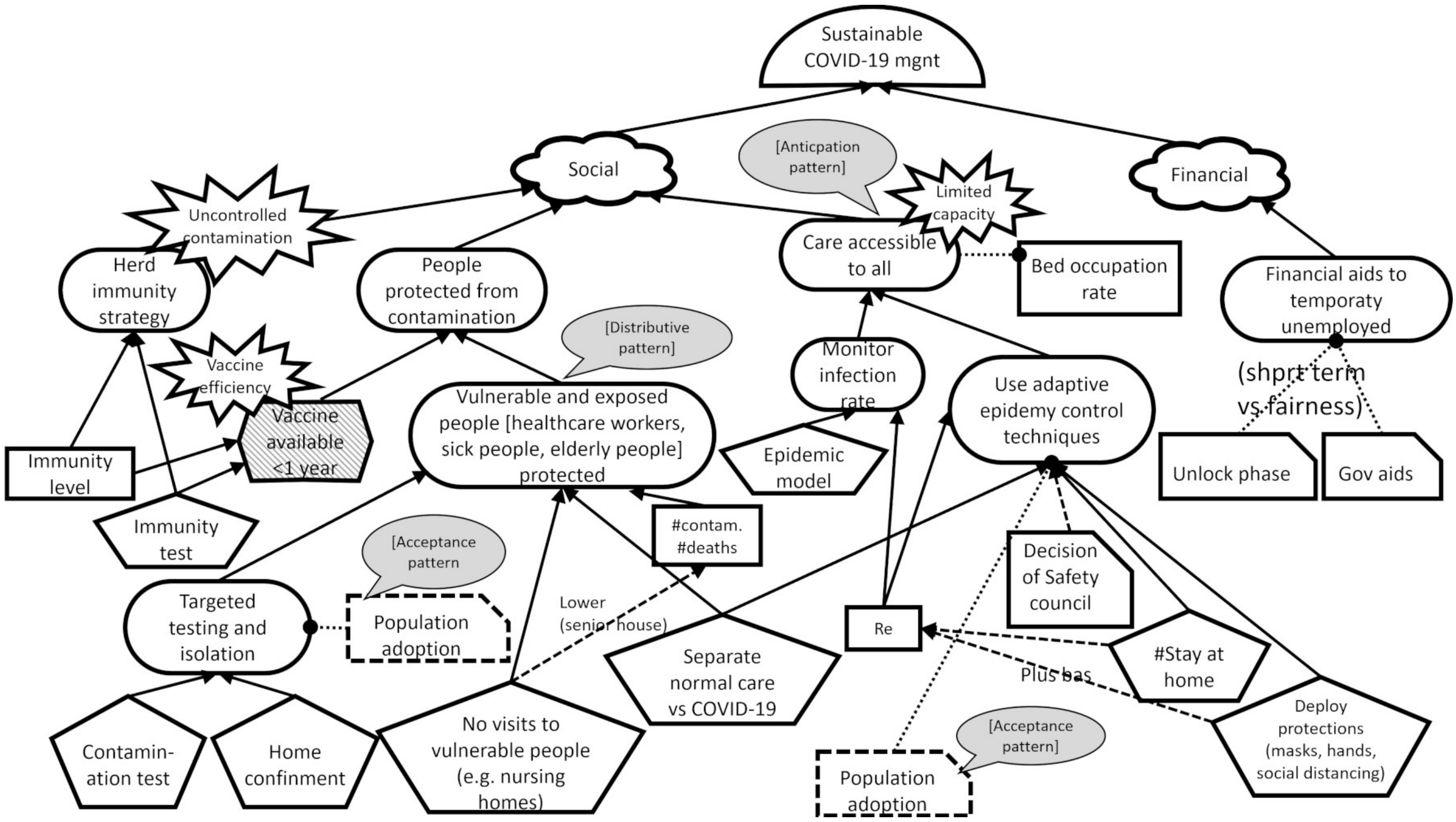

**Figure 13:** Partial analysis of the fairness of the containment phase of the COVID-19 pandemic

On the left-hand side of the diagram, the obstacles to herd immunity (at risk) and vaccination (long-term) are first identified, before focusing on the set of measures currently widely applied. Decoding the breakdown of the different values leads to the identification of the following potentially applicable patterns :

— The protection of the most vulnerable members is a pattern of distributive justice geared towards

these people. Activities to protect them are identified.

— The success of the approach depends on the acceptance of the people as underlined in the official communication calling for reciprocal responsibility of citizens and fairness [Bid20] [IC20]. This is captured by an assumption that can be developed beforehand in our model using our rule acceptance pattern. Note in this case that the acceptance of strict restrictions can be supported by human sciences such as psychology, by making explicit a goal to adhere to and by giving visibility on its realisation (see transparency pattern).

— Equitable access to care means that the system must operate within the limits of its resources and therefore anticipate overload. This means controlling the pandemic by flattening the contamination curve, which is now measured by the famous Re index, measuring the average number of people contaminated by a positive person, and which must be kept below 1. This index is monitored and an epidemic model is used to assess how the system is coping with the crisis, based on a range of measures (hygiene, masks, social distancing, lockdowns, etc.) which are continually evaluated and adapted, taking into account the latency of the disease of around 2 weeks.

— Compliance with these measures, particularly the confinement of people who are positive or at specific risk, requires a pattern of acceptance of the rules to ensure that they are adhered to.

### 4.2. *Pandemic COVID-19 - Vaccination phase(s)*

Since the end of 2020, the vaccination phase has been deployed. Its implementation has also revealed some interesting patterns, as shown in Figure 14. This diagram exploits the possibilities for refinement using the anchor point in the model shown in Figure 13. All the actions required for a complete and robust implementation of the vaccination are not detailed because they would require further decomposition : we limit ourselves to the main actions that make the approach comprehensible.

The value broken down here is the vaccination of the population. This was envisaged after a minimum period of one year, which was achieved thanks to the colossal research efforts implemented worldwide and which enabled several candidate vaccines to be explored in parallel. This point is captured by an assumption that has actually been implemented. Vaccination itself can therefore be based on the existence of a panoply of vaccines (leaving aside the authorisation phases). It follows a decomposition made up of the following milestones :

— The supply of vaccines is subject to problems of fairness, which are highly problematic in a multilateral approach where each country would have to try to source its own supplies from one or more pharmaceutical companies : countries would then be in competition with each other, with prices rising in favour of the richest, and there would be an increased risk of excessive orders to secure a minimum supply, and even risks of abuse, as was the case with the supply of masks, which followed such a pattern and even saw countries stealing shipments from other countries. In response to these dangers, a centralised approach based on a coordination framework is preferable. It has been implemented with European coordination, even if there are still aspects of competition between international blocs that are not, however, positioned entirely on the same vaccines. The element leading to the identification of the pattern in this case is the analysis of obstacles.

— The vaccination strategy, like the population protection strategy in the first phase, aims to protect the most vulnerable and exposed people. This strategy therefore follows the same pattern of distributive justice in determining vaccination priorities.

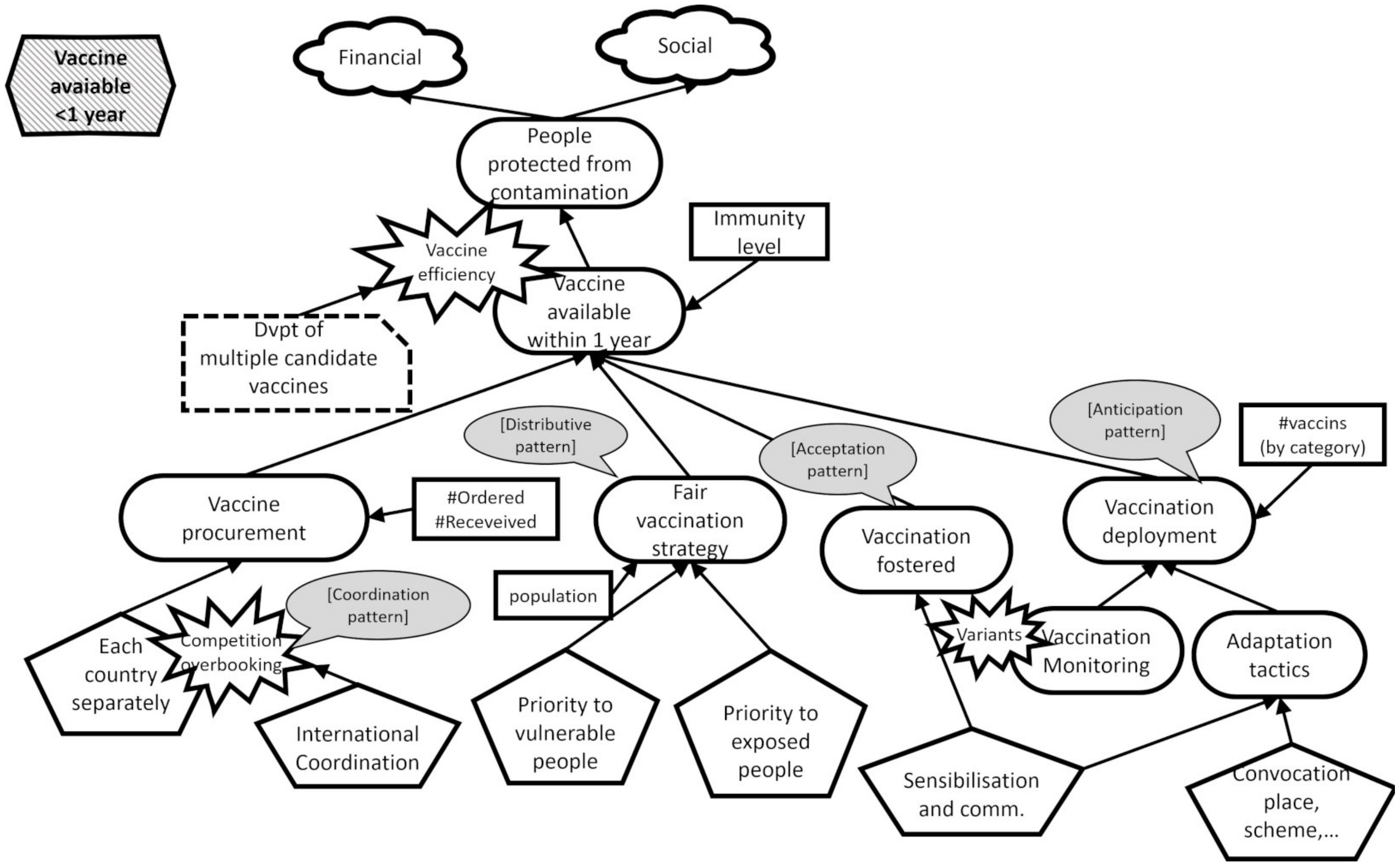

**Figure 14:** Partial analysis of the fairness of the vaccination phase of the COVID-19 pandemic

— Vaccination is not compulsory : no country seems to have imposed it, preferring to rely on people's support. It is therefore necessary to implement a strategy to ensure that people adhere to it, in particular by setting up awareness and communication campaigns as identified in this strategy.

— Finally, vaccination must be carried out at a sustained pace to reduce the risk of infection as quickly as possible and achieve a level of herd immunity. The process must be adaptive to take account of changing realities on the ground. For example, the appearance of variants means that the vaccination schedule needs to be adapted (spacing of doses, types of vaccine). The notification system and vaccination resources must be kept operational at the required rate to ensure that everyone notified receives their dose on the day of their appointment. In this case, too, obstacles are removed with a pattern that enables them to be anticipated as effectively as possible.

### 4.3. *Medical and social monitoring of early childhood*

The mission of the Office de la Naissance et de l'Enfance (ONE) is to approve, subsidise, organise, support, monitor and evaluate childcare and antenatal clinics, as well as out-of-home care for children under the age of 12. The organisation involves more than 1,800 workers, 1,000 doctors and 4,000 volunteers. It relies on around 400 consultations of various types. Its guiding principle is fairness : universality, non-discrimination and accessibility for all.

Figure 15 illustrates our analysis based on the values of universality and accessible, free care.

— the accessibility pattern is used here. This has not been detailed because it uses standard RE decomposition mechanisms. Accessibility for pregnant women and for young mothers with their babies is considered separately. In the first case, the proposed facilities are neighbourhood or hospital consul-tations. In the second case, the breakdown is based on three different locations. Firstly, the home, for the first few days, to avoid taking the baby out, but also to get a feel for the family environment. This point also enables us to detect social situations requiring closer monitoring to give all babies an equal chance. Thereafter, monitoring is carried out in one of the 400 clinics available, which should be close to the family. Finally, in order not to disadvantage certain sparsely populated areas, a mobile consultation service (via health coach tours) has also been introduced.

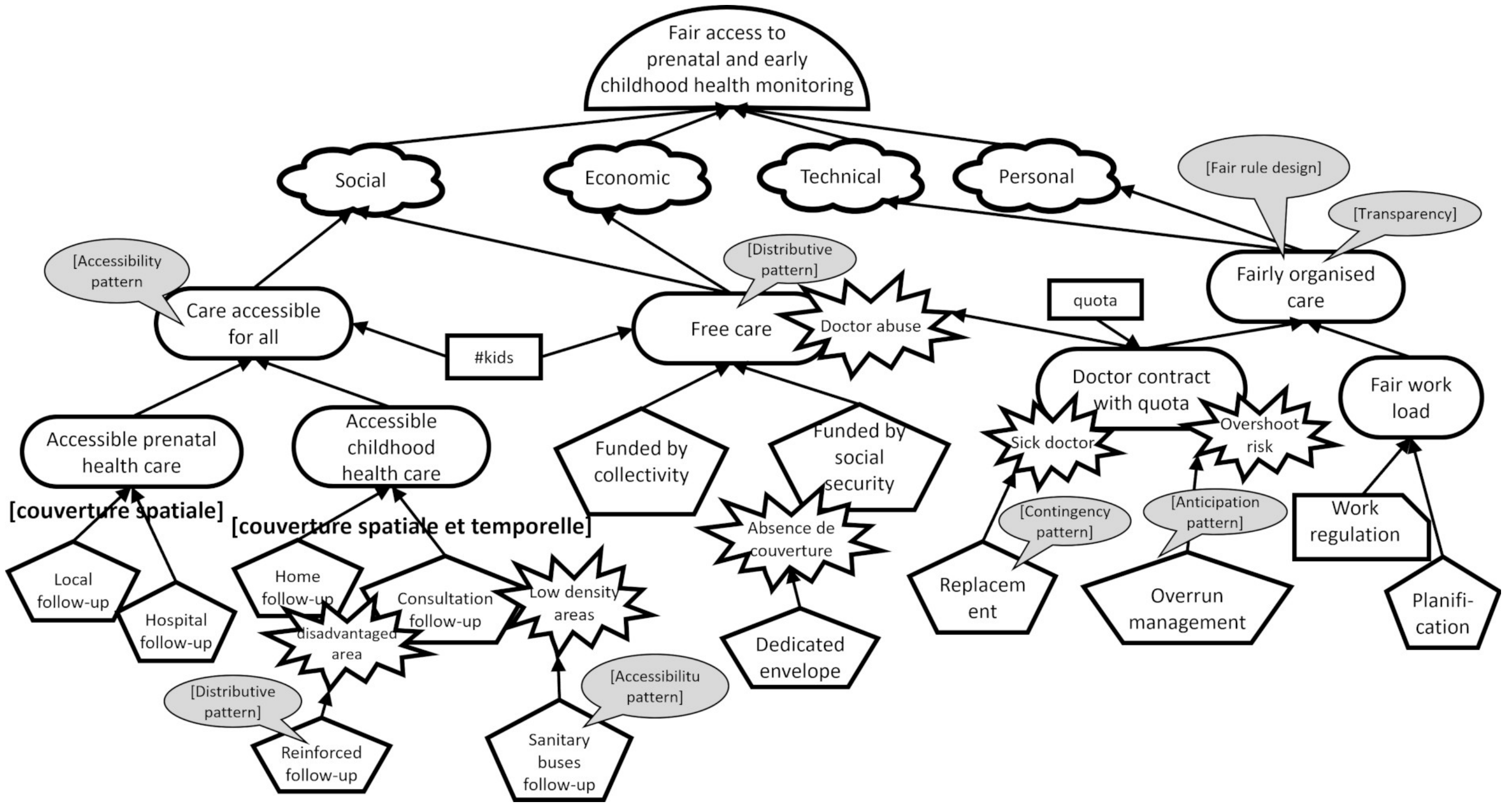

**Figure 15:** Partial analysis of the Office de la Naissance et de l'Enfance case

— The principle of distributive justice is implemented in the form of free healthcare. This puts all families on an equal footing. A more detailed analysis of the mechanism shows that it is based on subsidies to the ONE but also on social security cover. On this last point, however, there may be an obstacle for disadvantaged families who do not have such cover. This is covered by a specific budget allocation.

— The organisation of care falls within the scope of work and its organisation complies with the regulations in force. This requires complex planning, particularly by professionals in the health and social sector, which is carried out with a concern for transparency. It should be noted that in this field, the majority of women are employed.

— As far as doctors are concerned, they are under external contract with a quota of consultation hours based on estimates. Two obstacles can affect the availability of doctors for consultations : their absence or the exhaustion of their quota, which can be linked to attendance being higher than estimated. These cases are respectively dealt with by two of our patterns : absence by a contingency mechanism allowing one resource to be replaced by another, and the anticipation pattern which allows a quota to be extended by means of an amendment to the contract. Transparency is also important.

Figure 16 focuses on the value relating to the equitable organisation of doctors' consultations. The patterns mentioned in the previous paragraph are instantiated in a specific diagram devoted to the value of equitably organised care in order to avoid too many concepts complicating the initial figure.

— The first pattern instantiated is the one relating to the anticipation of quota violations. The generic terminology describing the pattern is adapted to the context : the load indicator here is the percentage of quota consumed. The IS can automatically detect this because it has all the information from actual and planning to check that the annual estimate (which is assumed to be realistic) is not in danger of being breached. If this is the case, the IS can also play a part in resolving the problem : firstly by calculating the additional quota that needs to be set aside, and secondly by helping to identify which doctor is best qualified to carry it out. The first candidate is the doctor currently assigned if he is not overloaded, but if not, another doctor known to the IS can be identified on the basis of his level of availability and the geographical area. There then remains a manual validation phase and production of a rider to the selected doctor. In this case, we can see that the documented pattern is largely respected, but tailored to the context.

— The second pattern instantiated concerns the need for transparency of operations. This pattern covers several complementary aspects : audit, software verification and publication of proof. In our case, only the last point was deemed relevant and sufficient, through the publication of quota information and calculation rules, including in particular the average time that we estimate a doctor should spend per child in consultation. In the end, the diagram reveals in grey a series of requirements placed on the IS so that it contributes to making the socio-technical system complete, fairer and more sustainable.

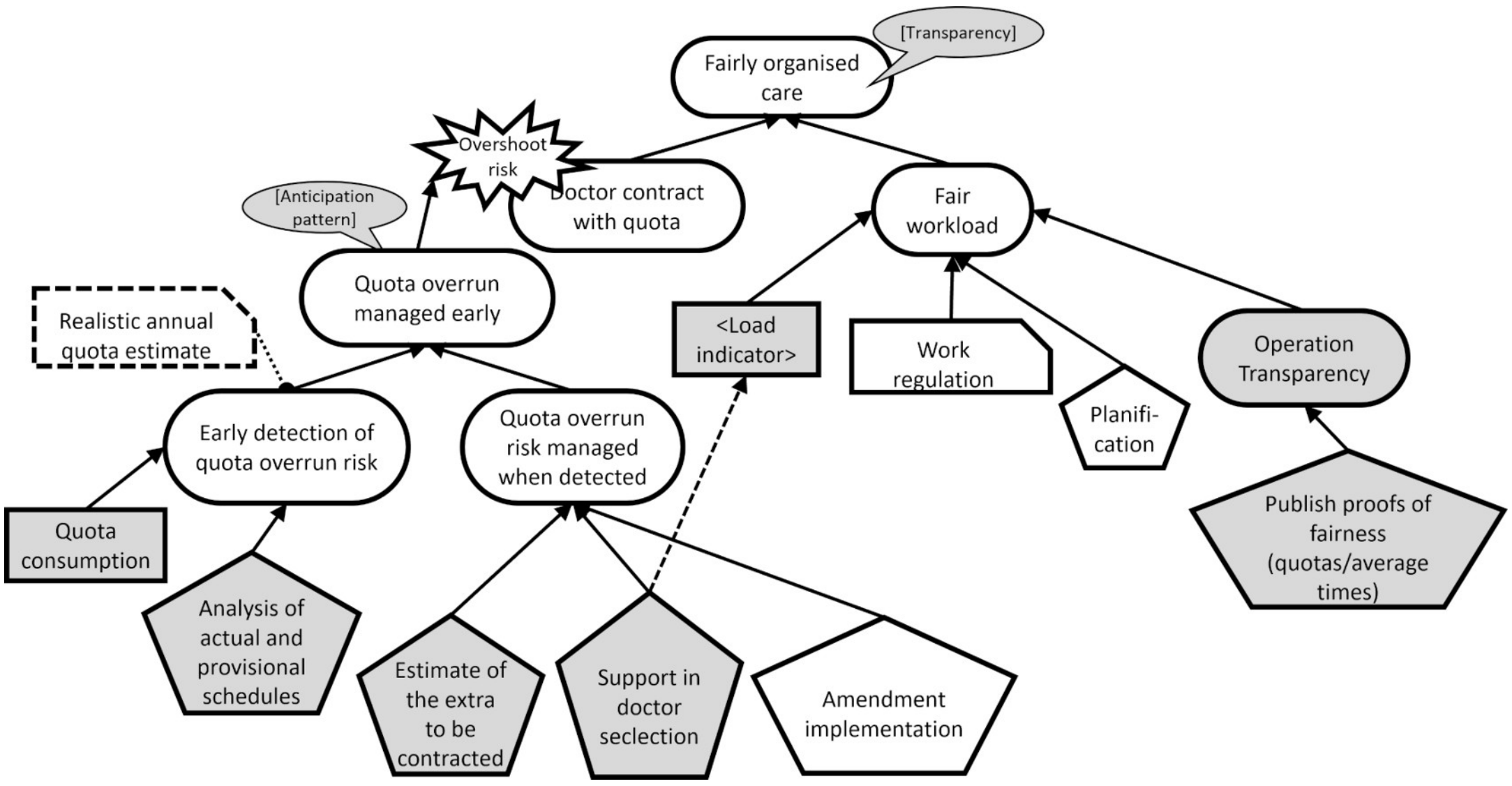

**Figure 16:** Instantiation of fairness patterns for the equitable organisation of care

## 5. Discussion

The creation of the catalogue of fairness patterns, followed by the validation process, highlighted a series of interesting points for discussion relating to the notations, the proposed extensions, the structuring of the catalogue and the way in which it could be enriched and applied more widely to the field of sustainability.

Concerning the language itself, it is centred on the key concept of value, expressed as a moral or natural good and perceived as the expression of a specific dimension [Pen13]. In essence, it shares similarities with the concept of purpose which captures, at different levels of abstraction, the objectives that the system under consideration should achieve [Lam01]. Values can also resemble non-functional requirements (e.g. maintainability, long-term use). In order to align the proposed meta-model with those of goal-oriented IE, our meta-model has grouped together the notions of goal, value and assumption by means of a specialisation. This makes it possible to associate obstacles with them. The links associating goals, values (via a dimension) and sub-values are similar to the refinement form « AND » of goals. The « OR » refinement form is not currently covered and is discussed below. Note that indicators could also be decomposed into more precise sub-indicators.

Regarding our extensions to the notations initially proposed [PF13], the introduction of the concept of obstacle has proved particularly rich, as illustrated in our validation cases. The presence of obstacles makes it possible to identify patterns in order to deal with them. For example, when fairness is desired, it can highlight situations that fall outside the framework of defined treatments and manage them specifically. This makes it possible to move away from a purely egalitarian framework and put in place identified strategies to protect the most vulnerable. Another example is the violation anticipation pattern (much more general) which can easily be connected to a mechanism for monitoring a key indicator indicating that the system is no longer capable of guaranteeing a certain form of fairness and reacting accordingly.

Our second extension concerns modularity. While it has been applied to pattern capture, it has only been used occasionally for our cases, which present relatively complex diagrams. The only use concerns the link between the two COVID processing phases. However, this is a deliberate choice of presentation. Our diagrams could easily have been broken down into a hierarchy of simpler diagrams, limiting their complexity to around fifteen concepts in order to remain within the rules of readability. In this respect, it should be noted that mapping points to a pattern are interesting candidates for referring to a diagram that specifies the pattern. This avoids overloading the initial diagram and keeps the essence of the pattern intact within a diagram instead of unfolding it in a diagram where it will be combined with other concepts.

Other notational extensions could also have been considered. In particular, we could have integrated the notion of alternative, which is present in RE languages such as KAOS and i*. This would have made it possible to reason about and evaluate multiple alternatives, at the cost of making the meta-model more complex. Also, the notions of agent or resource could have been made more explicit. We have not considered this at this stage, as the need did not emerge during the work carried out and out of a concern not to make the meta-model unnecessarily complex.

In terms of the catalogue structure, we initially opted for a hierarchical structure, which we changed to a cycle concept, with the possibility of positioning our patterns more appropriately in relation to the different stages identified. The main reasons for this are linked to the main pattern, which expressed such a process in a linear fashion and did not make explicit its iterative nature, or the need for governance. It also turned out that certain patterns often contributed to more than one stage. The possibility of positioning them in two dimensions offers an additional degree of freedom that is worth exploiting.

This work demonstrates the value of design patterns. These have been identified in several published sources and then applied on multiple occasions, quite systematically on various parts of our validation cases. Although our approach is spread over several years and includes validation on cases carried out in different contexts, this type of work requires a continuous process of enrichment and improvement, ideally carried out by an entire community. To date, it has only been tested by a small group of researchers close to the authors, who do not yet have a mechanism for wider dissemination or for gathering feedback, which is its main limitation. At this stage, while it remains difficult to estimate the level of quality or usefulness of the catalogue, we can nevertheless highlight factors on which feedback is positive : ease of learning, awareness of value-based reasoning and the clarification of a problem often left implicit or relegated to the level of secondary non-functional requirements. With regard to the COVID-19 case study, the patterns for vaccination and containment were proposed at the start of the phases concerned, without benefiting from a period of hindsight. The initial models proved to be very stable and have been reproduced here virtually unchanged. This indicates the high quality of the methodology and the catalogue.

The proposed approach can also deal with aspects other than the fairness dimension. A natural approach is to focus on the notion of resource, characterising its nature (renewable, reusable, disposable, etc.), the transparency of its environmental footprint, and the control of its use (alternatives, avoidance, optimisation, reuse, etc.). Some of these patterns have been documented in [HC15] and are directly linked to our anticipation pattern, which aims to avoid resource scarcity. It is also interesting to apply it to a wider field in order to evaluate the proposed extension of the meta-model and improve it further if necessary, while keeping it simple and focused on sustainability reasoning. More precise goal-oriented languages can come into play to support a subsequent and more detailed design analysis of the system under consideration, including the various components of its IS.

## 6. Conclusion and Perspectives

With this article, we contributed to the identification and documentation sustainability patterns addressing fairness for socio-technical systems. For this purpose, we proposed and tested a methodology for modelling and structuring patterns based on an existing modelling framework that we extended to provide more reasoning capabilities. We relied on a solid set of cases from the literature and from personal projects to support both the activities of pattern discovery and validation.

At this stage, our work has revealed real added value in terms of organising and re-using knowledge on the issue of fairness. This article remains a snapshot of a catalogue of fairness patterns that is bound to evolve. However, it already offers a good coverage of the various categories as evaluated on a continuous improvement cycle. Its usefulness has been confirmed by the validations carried out so far. The goal-based refinement and obstacle avoidance mechanisms borrowed from the Requirements Engineering domain also enable to use proven techniques to provide quality guarantees.

Based on these results from the specific field of fairness, we are now considering several avenues of research. At a practical level, it is important to assess the usability of the proposed ratings for the target audience, and also to provide appropriate tools for modelling, consulting patterns and ensuring that the catalogue is shared and enriched. Starting from the gained experience, a natural question is the wider exploration of sustainability using a similar approach, or even reconsidering the notion of non-functional requirements from another angle, since sustainability is linked in particular to maintainability, portability and usability. Along with security and dependability, it forms the core of the requirements of our time [Pen+14]. Finally, the information system point of view also deserves to be investigated in more detail in several directions. Patterns more specific to information system could be identified to make the link with software eco-design patterns.

## Acknowledgement

This work was partly funded by the IDEES CO-INNOVATION prokect of the Walloon Region (grant no ETR121200001379).